# Localization and migration of phase singularities in the edge-diffracted optical-vortex beams

Aleksandr Bekshaev[1*], Aleksey Chernykh[2], Anna Khoroshun[2], Lidiya Mikhaylovskaya[1]

[1]*Odessa I.I. Mechnikov National University, Dvorianska 2, 65082 Odessa, Ukraine*
[2]*East Ukrainian National University, Pr. Radiansky, 59-A, Severodonetsk, Ukraine*
[*]*Corresponding author: bekshaev@onu.edu.ua*

When a circularly-symmetric light beam with optical vortex (OV) diffracts at an opaque screen with the sharp edge, the OV core is displaced from the beam axis and, in case of the $m$-charged incident OV, decomposed into $|m|$ single-charged ones. By means of numerical simulations and based on examples of incident beams with topological charges $|m|$ =1, 2, 3 we show that, while the screen edge monotonously advances towards the beam axis, the OVs in the diffracted beam cross section move away from the incident beam axis along spiral-like trajectories. The trajectories contain fine structure details that reflect the nature and peculiar spatial configuration of the diffracting beam. For the Kummer beams' diffraction, the trajectories contain self-crossings and regions of "backward" rotation (loops); in case of Laguerre-Gaussian beams, the trajectories are smoother. The numerical results are supported by analytical approximations and conform with experiment The general shape of the trajectories and their local behavior show high sensitivity to the diffraction conditions (spatial structure of the diffracting beam, its disposition with respect to the screen edge, etc.), which can be used in diverse metrological applications.

## 1. Introduction

Light beams with optical vortices (OV) [1–4] attract growing attention during the past decades. Besides the rich and impressive physical contents, including the phase singularities, internal energy circulation, specific features of the linear and angular momentum distributions [4–6], such beams offer a wide range of perspective applications in the micromanipulation techniques [7–10], information transfer and processing [11,12], sensitivity and resolution enhancement in optical observations and measurements [13–19].

   Among diverse manifestations of the specific "circulational" properties of optical vortices, an important place belongs to the edge diffraction phenomena [20–35]. A series of experimental and theoretical researches has demonstrated that diffraction "reveals" the internal energy circulation, normally "hidden" in circular OV beams, and induces the essential transformation of their phase profile. In particular, even at weak screening (when the opaque obstacle covers only a far periphery of the beam cross section and induces no visible deformation of the intensity profile), the axial OV of the incident beam changes its position and, if its absolute topological charge $|m| > 1$, splits into a set of $|m|$ single-charged "secondary" vortices. As a result, a complicated pattern of singular points ("singular skeleton") is formed in the diffracted beam cross section that is highly sensitive to the diffraction conditions, especially to the screen edge position with respect to the incident beam axis.

A detailed study of these singularities and their evolution with the diffracted beam propagation was recently undertaken [32–34]; however, the theoretical discourse of these works was restricted to situations where the incident OV beam is described by the Laguerre-Gaussian (LG) model. This is the standard OV beam model but in practice, other sorts of OV beams are more appropriate. Frequently, experimental OV beams are formed from the usual Gaussian laser beam with the help of spiral phase plates or holographic gratings with groove bifurcation ("fork" holograms), which produce the OV beams of the Kummer family, or hypergeometric-Gaussian beams [36–38]. It is the use of Kummer beams that made possible recent experimental investigation of the singular skeleton in the diffracted OV beams and demonstration of the spiral-like motion of the secondary OVs when the screen edge performs a translational motion within the incident beam cross section [35]. In this work, we present the detailed numerical and theoretical analysis of this motion, disclose its peculiar features in the diffracted Kummer beams in comparison with their LG counterparts and discuss its relations to the physical nature of the OV beams.

## 2. Incident beam configurations and the diffraction conditions

We consider the usual scheme for experimental observation of the optical wave diffraction [35] illustrated in figure 1. The monochromatic laser beam (in our conditions, the wavelength is $\lambda =$ 632.8 nm), after standard filtering, expanding and shaping procedure (not shown in figure 1) forms the incident circular OV beam with the topological charge $m$ that experiences diffraction on the screen S edge. The incident beam propagates along the axis $z$, and, as usual in paraxial beams [3,39], the electric field distribution is described by expression $u_i(x,y,z)\exp(ikz)$ where $k = 2\pi/\lambda$ and $u_i(x,y,z)$ is the beam complex amplitude slowly varying on the wavelength scale.

Let in the screen plane the incident beam complex amplitude equals to $u_i = u_a(x_a, y_a)$; then the diffracted beam complex amplitude in an arbitrary observation plane situated at the distance $z$ behind the screen is determined by the Fresnel integral

$$u(x,y,z) = \frac{k}{2\pi i z} \int_{-\infty}^{\infty} dy_a \int_{-\infty}^{a} dx_a \, u_a(x_a, y_a) \exp\left\{\frac{ik}{2z}\left[(x-x_a)^2 + (y-y_a)^2\right]\right\}. \quad (1)$$

This equation was used many times for analysis of the OV beam diffraction (see, e.g., [32–35]).

Frequently in practice, the OV beam is formed with the help of a special element (VG in figure 1(a)) – a helical phase plate or a diffraction grating with groove bifurcation ("fork" hologram), which introduces the phase singularity into an initially regular (Gaussian) beam. The scheme of figure 1(a) exactly corresponds to the experimental conditions of [35]. In this case, the beam incident onto the screen belongs to the class of Kummer beams [37], $u_a(x_a, y_a) = u^K(x_a, y_a, z_h)$, described by the complex amplitude distribution

$$u^K(x_a, y_a, z_h) = \frac{z_{he}}{z_h}\sqrt{\frac{\pi}{2}}(-i)^{|m|+1}\exp\left[\frac{ik}{2z_h}(x_a^2 + y_a^2) + im\phi_a\right]\frac{z_R}{z_{he} - iz_R}e^{-A}\sqrt{A}\left[I_{\frac{|m|-1}{2}}(A) - I_{\frac{|m|+1}{2}}(A)\right] \quad (2)$$

where $z_h$ is the distance from the hologram to the screen (see figure 1(a)), $\phi_a = \arctan(y_a/x_a)$ is the azimuth (polar angle) in the screen plane, $I_v$ denotes the modified Bessel function [40],

$$A = \frac{1}{4}\frac{z_R}{z_{he} - iz_R}\left[\frac{k}{z_{he}}(x_{ae}^2 + y_{ae}^2)\right], \quad z_R = kb^2, \quad (3)$$

$$z_{he} = \frac{z_h}{1 + z_h/R}; \quad x_{ae} = x_a\frac{z_{he}}{z_h}; \quad y_{ae} = y_a\frac{z_{he}}{z_h}, \quad (4)$$

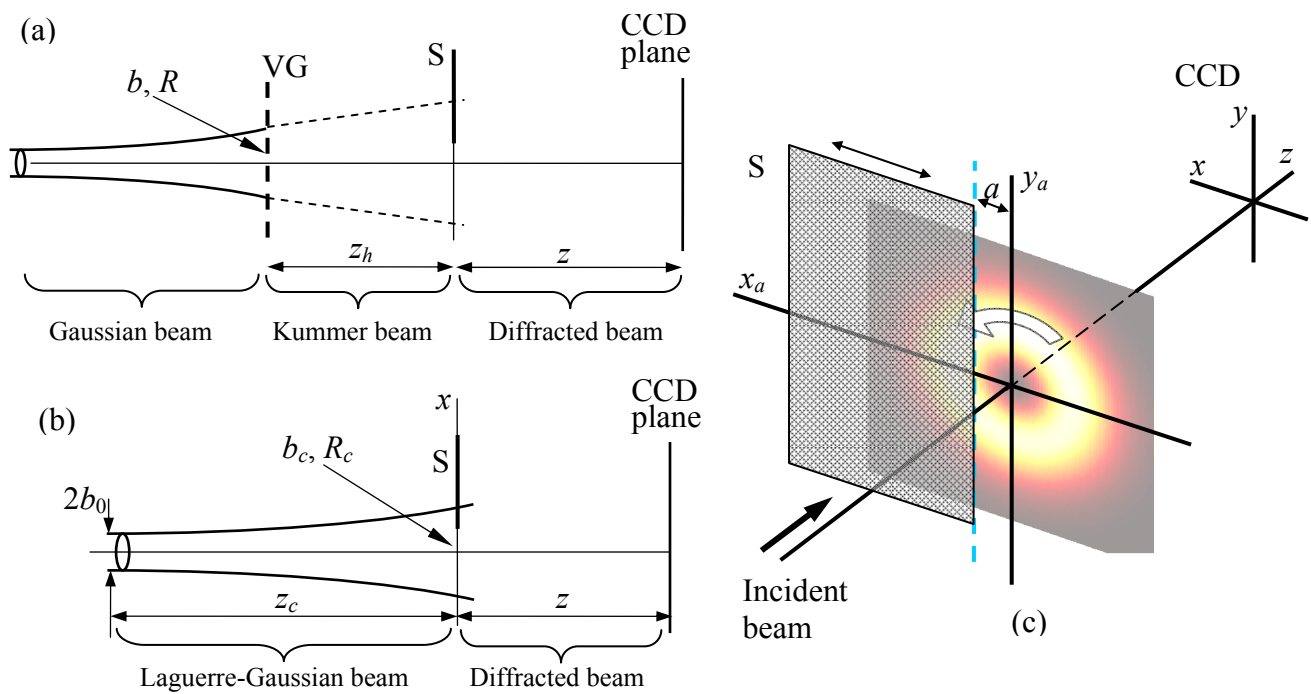

**Figure 1**. Scheme of (a) formation and diffraction of the incident Kummer beam and (b) diffraction of the incident LG beam; (c) magnified view of the beam screening and the involved coordinate frames. VG is the OV-generating element, S is the diffraction obstacle (opaque screen with the edge parallel to axis *y* and whose position along axis *x* is adjustable), the diffraction pattern is registered in the observation plane by means of the CCD camera. Further explanations see in text.

$b$ being the Gaussian beam radius at the grating plane VG, see figure 1(a). Equations (2) – (4) are written with allowance for the non-planar wavefront of the initial Gaussian beam [32,37], and $R$ is the radius of the wavefront curvature. Note that equation for $z_R$ in (3) formally coincides with the Raleigh range definition but looses the corresponding physical meaning since $b$ is no longer associated with the beam waist.

In experiments of [35], the following conditions for the incident beam were realized:
$$b = 0.232 \text{ mm}; \quad R = 54 \text{ cm}, \quad z_h = 11 \text{ cm}, \tag{5}$$
and, for convenience of references and illustrations, the data (5) will be used in the present analysis.

## 3. Simulation results: Kummer beams

In the course of simulations, we calculated the diffraction integral (1) that enables to obtain the amplitude and phase profiles of the diffracted beam in any cross section behind the screen. Examples of the diffraction beam profiles can be found elsewhere [32–35]; in this work, the calculated complex amplitude distributions were used for localization of the OV cores. The employed procedure includes manual identification of the amplitude zeros or the phase singularities within the calculated "maps" of certain areas within the beam cross section [33,34], which limits the accuracy of OV localization but permits clearly observing the general features of the OV migration and distribution. On the other hand, this procedure represents a sort of numerical experiment and models main difficulties and inaccuracies that may occur in a real experimental process.

In this paper, we present the results illustrating the OV migration in two fixed cross sections of the diffracted beam when the beam screening grows (screen edge moves from the far transverse periphery to the *z*-axis). Examples of the OV trajectories are shown in figure 2 and represent the refined and supplemented results of paper [35] where the favourable comparison with experimental data was also reported.

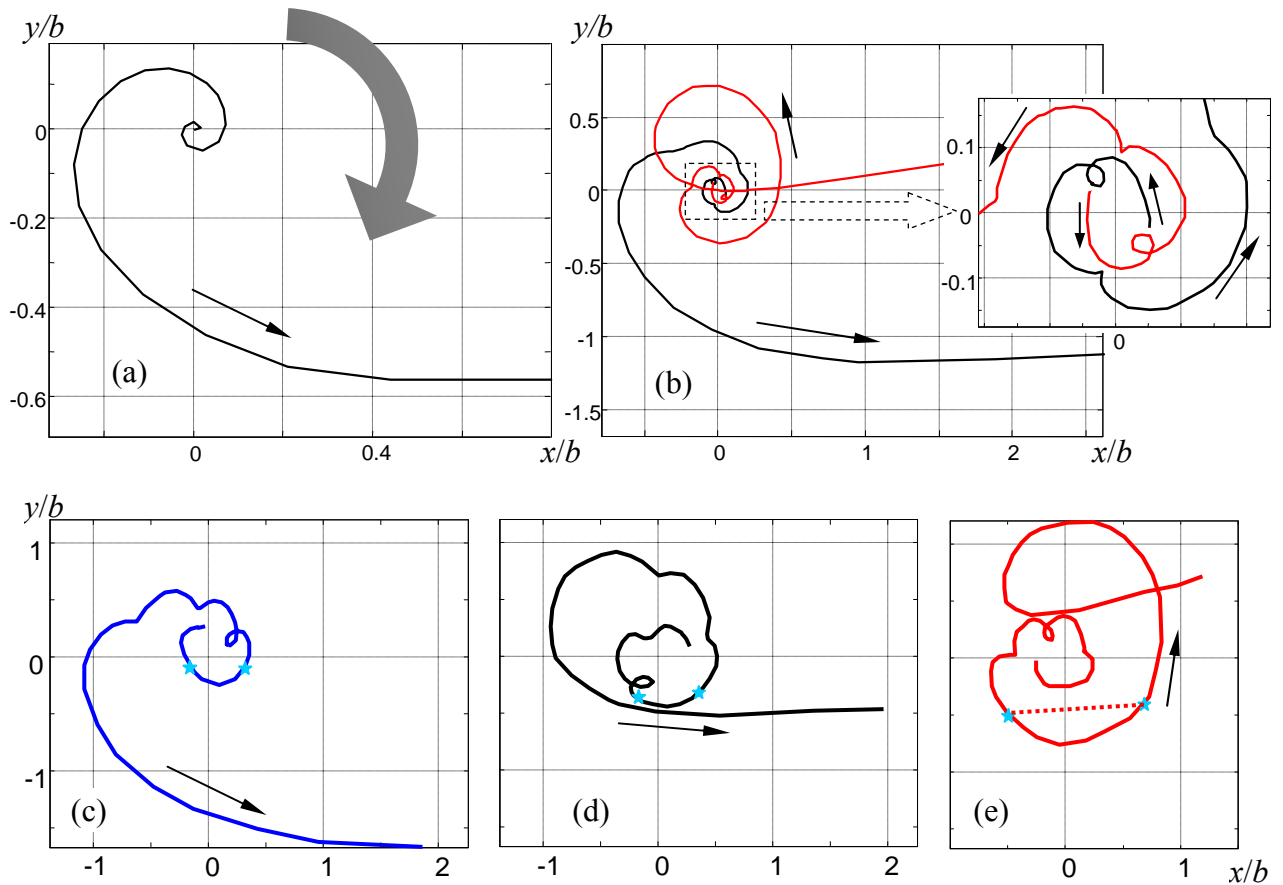

**Figure 2**. Trajectories described by the OV cores in the diffracted beam cross section $z = 30$ cm behind the screen when the screen edge moves towards the negative $x$ (see figure 1c), for incident Kummer beams with topological charge (a) $m = -1$, (b) $m = -2$ and (c)–(e) $m = -3$. The horizontal and vertical coordinates are in units of $b$ (5); large grey arrow shows the energy circulation in the incident beam (cf. figure 1c), small arrows show the direction of the OV motion when the screen edge approaches the incident beam axis; inset in the panel (b) shows a magnification of the near-zero region marked by dashed line. The dotted line between the cyan stars in panel (e) illustrates the OV "jump" (see the post-publication note, page 16).

For the single-charged input OV beam (figure 2(a)), the presence of the screen edge causes the OV core displacement from its initial position at the $z$-axis; with growing screening (decreasing $a$, see figure 1(c)), the OV core evolves along the spiral-like trajectory oppositely to the beam profile rotation, which, in turn, agrees with the internal energy circulation shown by the grey arrow. Ultimately, the OV disappears in the shadow region ($x > a$) [21–23,33,35]. When the screen edge moves away from the $z$-axis, the spiral-like OV trajectory makes (theoretically) an infinite number of rotations until it restores the initial axial position.

In case of the $m$-charged OV beam diffraction, the situation is complicated by at least two circumstances: (i) the axial OV decomposes into a set of $|m|$ secondary OVs that evolve separately and (ii) their spiral-like trajectories are much more intricate: with cusps, self-crossings and other singular points, which could not be reliably resolved in experiment [35]. This is illustrated by the examples in figures. 2(b)–(e); note that the special colour is given to each OV, and this convention is also kept in subsequent figures 3 – 5. For the 3-charged incident Kummer beam, the whole configuration is so complicated that the evolution of secondary OVs is shown in separate panels (c), (d) and (e). The pattern is especially knotty at small screening ($a \gg b$) when the OVs are slightly perturbed and situated near the nominal beam axis (see the inset in figure 2(b)). Of course, the OV trajectories, again, make infinite number of rotations when approaching the origin, which cannot be shown in figures 2(b)–(e).

Because of the complex character of the OV trajectories, their parametric representation in polar coordinates $r$, $\phi$ where

$$x = r\cos\phi, \quad y = r\sin\phi, \tag{6}$$

looks more informative; this representation is used in figures 3 – 5. Although the curves in figures 3 and 4 are not so spectacular as the explicit trajectories of figures 2, they offer additional information relating the rate of the OV motion caused by the uniform motion of the screen edge (in other words, the sensitivity of the OV position to the screen displacements). In this context, it is interesting the rapid growth of radial coordinates of the secondary OV near $a = 0$ in figures 3(a), (b); near $a = -0.5b$ and $a = 0.3b$ in figures 3(c), (d); and near $a = -0.7b$, $a = 0$ and $a = 0.6b$ in figures 3(e), (f). In figure 4(e), one can see remarkable azimuthal "boosts" of the blue curve near $a = 3.7b$, of the black curve near $a = 3.1b$ and of the red curve near $a = 2.3b$ that correspond to very fast evolution of the corresponding OVs between the points marked by cyan stars in figures 2(c)–(e) (see the post-publication note, page 16). Different OVs experience this "acceleration" at different $a$ but every time it occurs when the trajectory passes below the axis $z$ against the direction of the moving screen edge. Noteworthy, while one of the OVs quickly passes the "rapid" segment of its trajectory, the other ones perform the "normal" slow motion typical for other trajectory segments.

Many other details of the OV trajectories find their counterparts in behaviour of the curves in figures 3, 4. Generally, the spiral motion is characterized by monotonic growth of $\phi$; however, regions of the "backward" evolution of the azimuth are present in figures 4(c)–(f) that are associated with the trajectories' self-crossing ("loops") distinctly seen in figures 2(b)–(e). The π-jumps of the red curves at $a = 0$, seen in figures 4(c), (d) ($m = -2$), appear because, for diffraction of a circular OV beam with the even topological charge, the corresponding OV trajectory obligatory crosses the axis $z$ [34,35]. For the same reason, the red curves for the radial OV coordinate in figures 3(c), (d) touch the horizontal axis at $a = 0$ and possess cusps there (cf. the red trajectory in figure 2(b)). Notably, these cusps, typical for even $|m|$, can be treated as if the red curves experience "mirror reflection" from the horizontal axis: the curves can be smoothly continued to "negative" $r$, which correspond to the π-jumps of the azimuth mentioned above. In contrast, for the odd topological charge ($m = -3$), there are smooth minima of black and red curves in figures 3(e), (f).

The evolution of the OV positions with the diffracted beam propagation is not the subject of the present work; for the LG beam diffraction, it was studied elsewhere [33,34]. However, comparison of the right and left columns of figures 3, 4 clearly demonstrates the main propagation-induced effects: the OV cores, generally, move away from the $z$-axis (the vertical scales are higher in the right column of figure 3), the spiral motion becomes slower (for each curve in the right column of figure 4, the range of $\phi$ variation is smaller than for the same colour curve in the left column), and the small visually irregular oscillations ("ripples") in the $r$ and $\phi$ curves become smoother.

## 4. Discussion and interpretation of the results

In general, the considered patterns of the OV migration are complicated but their main feature – the spiral character of the OV core evolution upon monotonic variation of $a$ – admits simple physical explanation. To this purpose, we resort to the asymptotic expression for the diffracted beam amplitude (1) derived in Appendix

$$u(x,y,z) \simeq B_1 \left(\frac{r}{b}\right)^{|m|} \exp(im\phi) - \frac{D_1}{a^3} \exp\left[\frac{ika^2}{2}\left(\frac{1}{z_h} + \frac{1}{z}\right)\right] + \frac{D_2}{a} \exp\left[\frac{ika^2}{2}\left(\frac{1}{z_d} + \frac{1}{z}\right)\right] \tag{7}$$

where $z_d$ is defined in (A17),

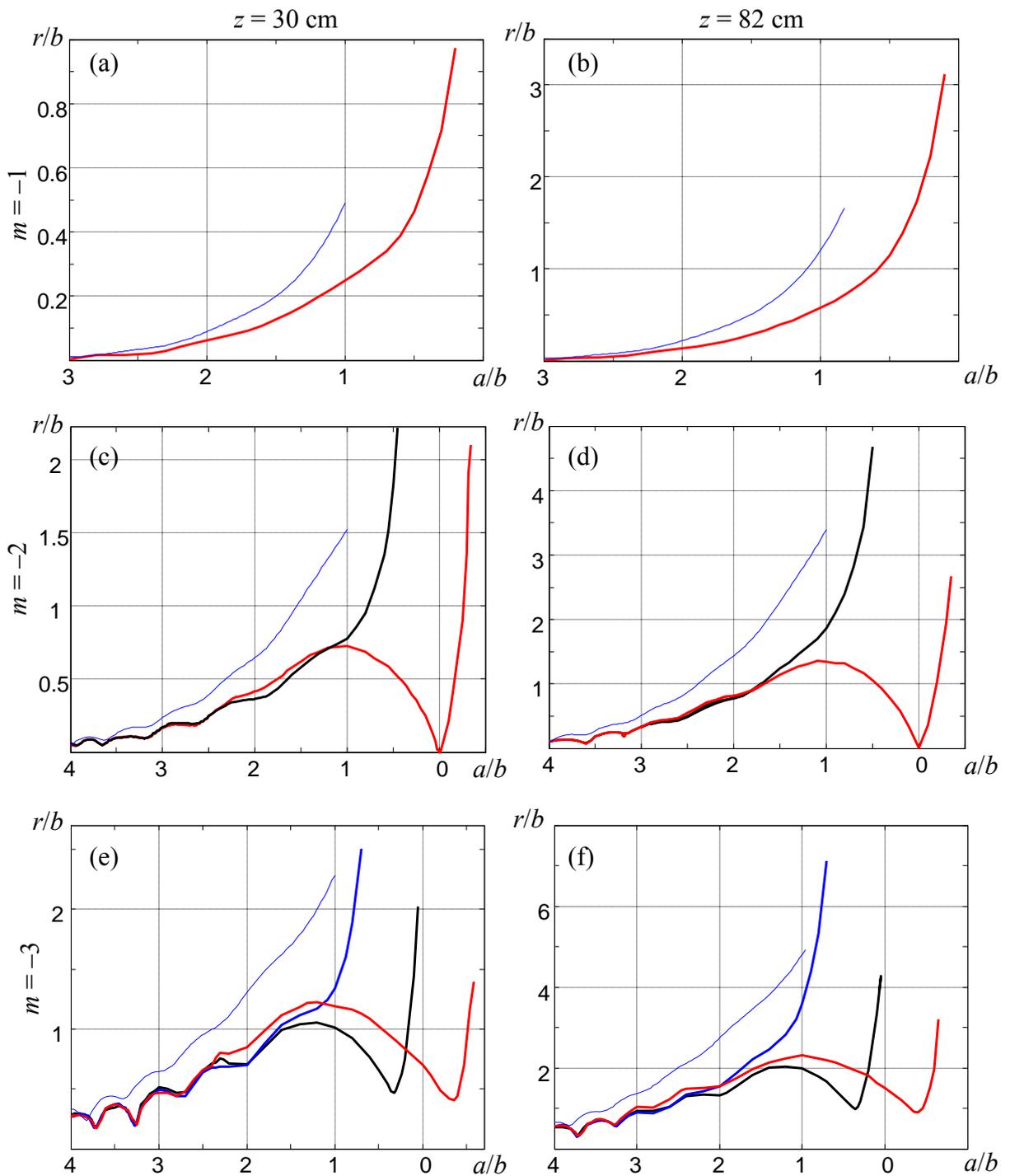

**Figure 3**. Radial coordinates (6) of the OV cores in the diffracted beam cross sections (marked above each column) for incident Kummer beams with topological charges (a,b) $m = -1$, (c,d) $m = -2$ and (e,f) $m = -3$. Thick curves illustrate the evolution of $r$ in units of $b$ (5) when the screen edge moves from large positive $x$ towards the beam axis and slightly further; inverse horizontal scale indicates the screen edge position $a$ in units of $b$ (5) (cf. figure 1(c)). Curves of different colors in panels (c) – (f) describe different secondary OVs that appear due to decomposition of the multicharged OV carried by the incident beam; thin curves represent the asymptotic approximation of (14), (15).

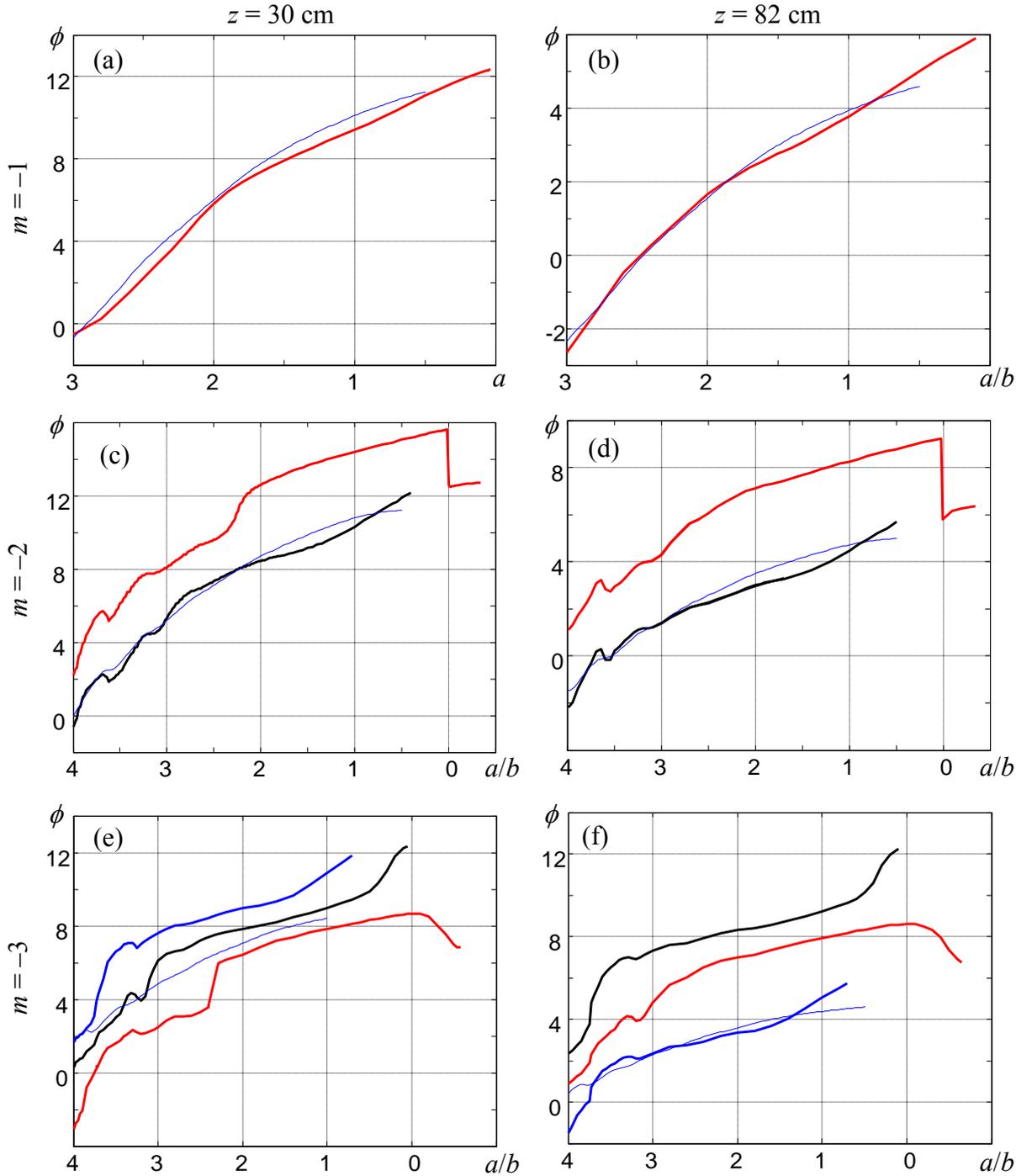

**Figure 4**. Angular polar coordinates (6) of the OV cores in the diffracted beam cross sections (marked above each column) for incident Kummer beams with topological charges (a,b) $m = -1$, (c,d) $m = -2$ and (e,f) $m = -3$. Thick curves illustrate the evolution of $\phi$ in radians when the screen edge moves from large positive $x$ towards the beam axis and slightly further; inverse horizontal scale indicates the screen edge position $a$ in units of $b$ (5) (cf. figure 1(c)). Curves of different colors in panels (c) – (f) describe different secondary OVs that appear due to decomposition of the multicharged OV carried by the incident beam; thin curves represent the asymptotic approximation of (14), (15).

$$B_1 = \sqrt{\frac{\pi}{2^{3|m|}}} \left(\frac{-iz_e}{z+z_h}\right)^{|m|+1} \frac{z_R^{|m|+1}}{z_e^{|m|/2}(z_e - iz_R)^{|m|/2+1}}, \tag{8}$$

$$D_1 = |m|(-i)^{|m|+1}\sqrt{\frac{i}{2\pi}}\frac{z_h}{z}\left[k\left(\frac{1}{z_h}+\frac{1}{z}\right)\right]^{-3/2}, \tag{9}$$

$$D_2 = -\frac{ik}{2^{2|m|}}\sqrt{\frac{i}{2\pi}}\frac{z_{he}}{zz_h}\frac{z_R^{2|m|+1}}{z_{he}^{|m|}(z_{he}-iz_R)^{|m|+1}}\left[k\left(\frac{1}{z_d}+\frac{1}{z}\right)\right]^{-3/2}. \tag{10}$$

Equations (7) – (10) are valid if $a \gg b$ and only near the axis, $x \ll b$, $y \ll b$ (see (A4)) but they express the physical essence of the diffraction process [39]. The first term of (7) describes the unperturbed incident beam while the role of the screen is accumulated by the second and third terms that describe a sort of the "edge wave" [39] which interferes with the incident beam, and thus the diffraction pattern is formed.

In a very simplified form, the edge wave can be considered as a cylindrical wave emitted by the screen edge. In the centre of the observation plane its complex electric field approximately amounts to

$$E_{\text{edge}} = D_0(a)\exp\left[ik\left(z+\frac{a^2}{2z}\right)\right] \tag{11}$$

(see figure 1(c)), with complex coefficient $D_0(a)$, $|D_0(a)|$ is a decreasing function of $a$, while the incident beam contribution can be expressed as

$$E_{\text{inc}} = B_0\left(\frac{r}{b}\right)^{|m|}\exp(im\phi)\exp(ikz) \tag{12}$$

(cf. the first term in the right-hand side of (7)), with certain complex constant $B_0$. The OV positions are determined by the condition $E_{\text{edge}} + E_{\text{inc}} = 0$, which entails

$$\frac{r}{b} = \left[\frac{|D_0(a)|}{|B_0|}\right]^{1/|m|}, \quad \phi = \frac{1}{m}\left[\arg D_0(a) - \arg B_0 + k\frac{a^2}{2z} + (2N-1)\pi\right], \quad N = 0, 1, \ldots |m|-1. \tag{13}$$

Equations (13) roughly describe the OV trajectories in the observation plane; in particular, they dictate that all the secondary OVs behave identically, with the only distinction in their azimuthal positions determined by $N$. However, the spiral-like motion of the point $(r, \phi)$ determined by (13), upon the monotonic variation of $a$, is clearly seen, as well as the infinite number of rotations with $a \to \infty$; additionally, the similarities in the evolutions of different OVs at high $a$ are also obvious in all the considered situations illustrated by figures 2 – 4.

Nevertheless, the simplified model of (11) – (13) gives nothing but a qualitative explanation. One can notice that the asymptotic (7) – (10) looks akin to (11) and (12) and, properly speaking, provides refinement and further elaboration of the same idea. As a result, it, indeed, supplies a reasonable quantitative approximation. By equating (7) to zero one immediately obtains the OV trajectories in the form similar to (13):

$$\frac{r}{b} = \left\{\left|\frac{D_1}{a^3} - \frac{D_2}{a}\exp\left[\frac{ika^2}{2}\left(\frac{1}{z_d}-\frac{1}{z_h}\right)\right]\right|\cdot|B_1|^{-1}\right\}^{1/|m|}, \tag{14}$$

$$\phi = \frac{1}{m}\arg\left[\frac{D_1}{a^3}\exp\left(\frac{ika^2}{2z_h}\right) - \frac{D_2}{a}\exp\left(\frac{ika^2}{2z_d}\right)\right] + \frac{ka^2}{2mz} - \frac{1}{m}\arg B_1 + \frac{2N}{m}\pi \tag{15}$$

(*N* is defined in the last equation (13)). It is this solution that is illustrated in figures 3, 4 by thin lines. One can see that the quality of the approximation is rather good for $a > 3b$ and for $|m| = 1$ but diminishes with increase of the topological charge of the incident OV beam. The approximation (14), (15) qualitatively reproduces even such fine details of the exact curves as the oscillatory variations of the OV radial coordinates in figures 3(c)–(f) for $a > 2b$, and local decreases of the $\phi$ curves for $a > 3b$ (figures 4(c)–(f)) responsible for self-crossings and loops in the trajectories (cf. figures 2(b)–(e)). However, relations (14) and (15) predict the identical behavior for each secondary OV trajectory that may differ only by the constant azimuthal offset described by the last term of (15) (that is why only one asymptotic curve is presented in each panel of figures 3 and 4).

What the asymptotic solution (14), (15) expectedly cannot explain, is the fine structure of apparently irregular oscillations noticeable in the thick curves of 2$^{nd}$ and 3$^{rd}$ rows of figure 3 and, especially, of 2$^{nd}$ and 3$^{rd}$ rows of figure 4. We suppose that this fine structure is associated with oscillations ("ripples") in the phase-amplitude profile of the incident beam which exists in any Kummer beam at moderate distances behind the OV-generating element due to influence of the embedded phase singularity [37,41]. To evaluate the role of this ripple structure, we have also perform the numerical analysis of the secondary OV behavior in case of the LG beam diffraction, where the ripple structure is surely absent.

For a LG beam with topological charge *m*, the complex amplitude distribution in the screen plane (see figure 1(c)) is determined by equation $u_a(x_a, y_a) = u^{LG}(x_a, y_a, z_c)$ where

$$u^{LG}(x_a, y_a, z_c) = \frac{(-i)^{|m|+1}}{\sqrt{|m|!}} \left(\frac{z_{R0}}{z_c - iz_{Rc}}\right)^{|m|+1} \left(\frac{x_a + i\sigma y_a}{b_0}\right)^{|m|} \exp\left(\frac{ik}{2} \frac{x_a^2 + y_a^2}{z_c - iz_{Rc}}\right), \qquad (16)$$

$\sigma = \mathrm{sgn}(m) = \pm 1$. Here $b_0$ is the Gaussian envelope waist radius, $z_c$ is the distance from the waist cross section to the screen plane (see figure 1(b)), and $z_{Rc} = kb_0^2$ is the corresponding Rayleigh length [39]; more usual "explicit" beam parameters, the current beam radius $b_c$ and the wavefront curvature radius $R_c$ in the screen plane, are related with these quantities by known equations

$$b_c^2 = b_0^2 \left(1 + \frac{z_c^2}{z_{Rc}^2}\right), \quad R_c = z_c + \frac{z_{Rc}^2}{z_c}. \qquad (17)$$

Numerical values of the LG beam parameters are chosen so that to facilitate a direct comparison with the above-considered Kummer beam diffraction, which requires the transverse profile of the incident LG beam at the screen plane to be in maximal possible similarity to the Kummer beam spatial profile of [35]. This non-rigorous requirement is satisfied if we accept $b_c = 0.23$ mm, $R_c = 57$ cm, which corresponds to the parameters of equation (16)

$$b_0 = 0.17 \text{ mm}, \quad z_{Rc} = 28.5 \text{ cm}, \quad z_c = 27 \text{ cm}. \qquad (18)$$

Behavior of the OV cores upon diffraction of the LG beam described by (16) – (18) was simulated on the base of equation (1) where expression (16) is substituted. Simultaneously, the analytic approximation analogous to that of (7) – (10) and based on (A1), (A8) and (A9), has been employed, which yields the following expression of the diffracted field in the observation plane:

$$u(x, y, z) \simeq Br^{|m|} \exp(im\phi) - Da^{|m|-1} \exp\left[\frac{ika^2}{2}\left(\frac{1}{z_c - iz_{Rc}} + \frac{1}{z}\right)\right] \qquad (19)$$

where

$$B = \frac{1}{(z + z_c - iz_{Rc})^{|m|+1}}, \quad D = \sqrt{\frac{i}{2\pi}} \frac{k}{z} (z_c - iz_{Rc})^{-|m|-1} \left[k\left(\frac{1}{z} + \frac{1}{z_c - iz_{Rc}}\right)\right]^{-3/2}. \qquad (20)$$

Hence, again, the resemblance with (11) – (13) is obvious, and explicit expressions for the OV cores' polar coordinates can be easily derived via equating (19) to zero. Accordingly,

$$r = \left\{ \frac{a^{|m|-1}}{|B|} \left| D \exp\left[ \frac{ika^2}{2} \left( \frac{1}{z_c - iz_{Rc}} \right) \right] \right| \right\}^{1/|m|}, \tag{21}$$

$$\phi = \frac{1}{m} \left\{ \arg\left[ D \exp\left( i \frac{ka^2}{2} \frac{1}{z_c - iz_{Rc}} \right) \right] - \arg B \right\} + \frac{ka^2}{2mz} + \frac{2N}{m}\pi. \tag{22}$$

The results are illustrated in figure 5 for the single cross section $z = 30$ cm, corresponding to left columns of fiugures 3 and 4. First of all, in this case the trajectories are, indeed, much smoother (light wrinkles that can be seen in the thick curves appear due to the limited accuracy of calculations and are negligible). The second distinction is that the radial displacements of the OV cores in figure 5(a), (b) are smaller and decrease with growing $a$ faster than in corresponding panels of figure 3(c), (e), which, evidently, can be ascribed to the lower divergence of the LG beam and to the rapid exponential decay of the edge-wave amplitude with growing $a$. At the same time, the thick curves of figure 5 fairly reproduce the general character of the OV trajectories, previously detected for the Kummer beams' diffraction, particularly, the cusp in the $r(a)$ dependences for $m = -2$ (compare figures 5(a) and 3(c)) and the azimuthal "boost" in case of $m = -3$ (red curves in figures 4(e) and 5(d)).

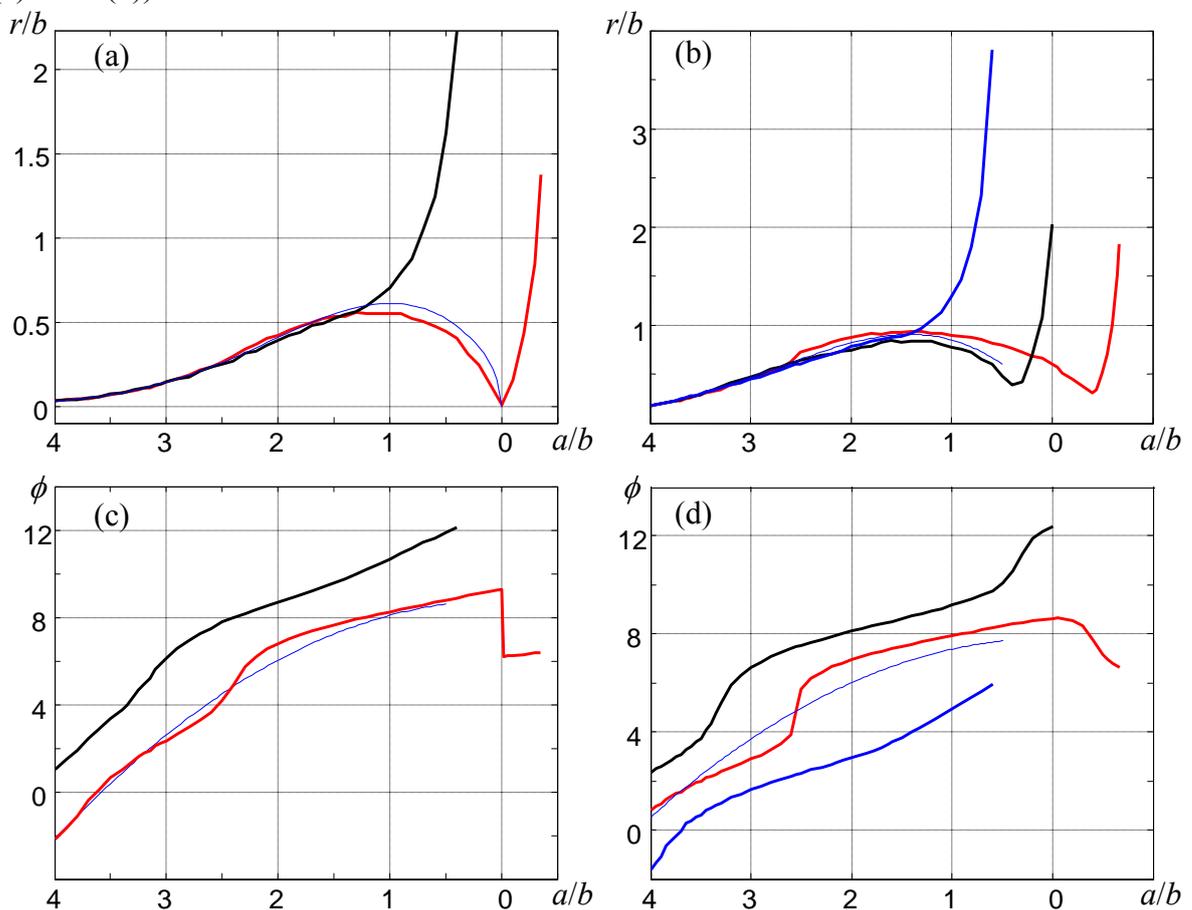

**Figure 5**. (a,b) Radial and (c,d) angular polar coordinates (6) of the OV cores in the diffracted beam cross section $z = 30$ cm for incident LG beams with topological charges (a,c) $m = -2$ and (b,d) $m = -3$. Thick curves illustrate the evolution of $r$ (in units of $b$ (5)) and $\phi$ in radians when the screen edge moves from large positive $x$ towards the beam axis and slightly further; inverse horizontal scale indicates the screen edge position $a$ in units of $b$ (5). Curves of different colors describe different secondary OVs that appear due to decomposition of the multicharged OV carried by the incident beam; thin curves represent the asymptotic approximation of (21), (22).

But the most impressive peculiarity of the LG beam diffraction is the rather high quality of the analytic approximation based on the asymptotic relations (20) – (22). In fact, the value $a/b \sim 4$ is too small to expect a good validity of asymptotic expressions, and the clear tendency of the thin curves in figure 3 to approach the exact results with growing $a$ is the best that could be expected. However, in figure 5(a), (b) the thin curves almost coincide with the exact radial evolution for one of the OV trajectories practically until $a = 0$. The agreement is not so complete for the OV angular coordinates (figure 5(c), (d)) where the thin curves miss the important "acceleration" of the OV rotation near $a/b = 2.4$. However, in this view, one can expect that the asymptotic formulas (21), (22) are not just an illustration of the physical model but possess their own independent value and can be used for practical calculations of the secondary OV positions in the diffracted LG beams.

## 5. Conclusion

In this paper, we have analyzed numerically and analytically a rather special problem associated with the OV diffraction: the spatial distribution and migration of the OV cores within the diffracted beam cross section. Despite the limited number of examples (Kummer and LG beams with topological charges $|m| \leq 3$) and restriction by two fixed cross section (which was dictated by adherence to the conditions of the paper [35] whose experimental results fairly conform with our theoretical analysis), the main findings of this paper reveal some general features of the diffracted OV beams which constitute common interest.

The spiral motion of the phase singularities upon the monotonous advance of the screen edge is a direct consequence of the helical nature of OV beams and can be considered as a distinct visual manifestation of the screw wavefront dislocation and the transverse energy circulation in the incident beam. Besides, the peculiar features of the OV trajectories, especially in case of milticharged OV diffraction, carry the fingerprints of the incident beam spatial structure and enable, e.g., to distinguish Kummer beams from the LG ones. Remarkably, the observed [35] and calculated trajectories of the diffracted beam singularities contain a rich set of fine details that need further comprehensive investigation and interpretation. It is reasonable to expect that this activity will elucidate the mechanisms of the OV diffraction and bring about novel efficient means for the OV diagnostics.

Interesting questions appear relating the sensitivity of the OV trajectories to the diffraction conditions. Figures 2 – 5 supply a lot of instances where a small change of the screen position $a$ with respect to the incident beam axis provokes huge transformations of the singular skeleton of the diffracted beam (for example, regions of rapid growth of the radial OV coordinates in figures 3(c)–(f) and 5(a), (b) or azimuthal "boosts" visible in red curves of figures 4(e) and 5(d)). One can expect that these features can be used for sensitive distant detection and measurement of the screen edge displacement with respect to the incident beam axis. On this base, practical measuring procedures can be devised which will contribute to further development of the OV metrology [15] and singularimetry [18,19] – intensively advancing research areas that employ unique metrological perspectives of the OVs.

Note that this development will require more detailed study of additional parameters of the secondary OVs, especially, their morphology [42–44] characterized by the orientation and the form-factor of equal intensity ellipses in the nearest vicinity of the OV core. Our preliminary observations show that the high sensitivity of an OV position to the screen displacement $a$ is often coupled with a strong OV anisotropy (in particular, this case is realized for the trajectory segments between the cyan stars in figures 2(c)–(e) as well as for regions of rapid growth of the OV radial coordinates in figures 3, 5). This anisotropy not only may affect the OV localization accuracy but its behavior with respect to $a$ can be an additional valuable source of information on the OV diffraction processes.

**Acknowledgments**

This work was supported, in part, by the Ministry of Science and Education of Ukraine, Project No. 531/15.

**Appendix**

For weak screening, $a \gg b$, $a \gg b_c$, equation (1) can be suitably represented in the form

$$u(x,y,z) = u^I(x,y,z) - \frac{k}{2\pi i z} \int_{-\infty}^{\infty} dy_a \int_a^{\infty} dx_a\, u_a(x_a, y_a) \exp\left\{\frac{ik}{2z}\left[(x-x_a)^2 + (y-y_a)^2\right]\right\} \tag{A1}$$

where $u^I(x,y,z)$ describes the complex amplitude distribution of an unperturbed beam (what had occurred in the observation plane if the screen were absent).

Let us start with considering the LG beam diffraction, then $u^I(x,y,z) = u^{LG}(x,y,z_c + z)$ (cf. equation (16)) while, in the integral in the right-hand side of (A1), the expression (16) enters immediately. Then, omitting the coordinate-independent multipliers of (16), the integral in (A1) acquires the form

$$\exp\left[\frac{ik}{2z}(x^2 + y^2)\right] \int_{-\infty}^{\infty} dy_a\, P(y_a, y, z_c - iz_{Rc}) \int_a^{\infty} dx_a\, (x_a + i\sigma y_a)^{|m|} P(x_a, x, z_c - iz_{Rc}) \tag{A2}$$

where

$$P(x_a, x, d) = \exp\left[\frac{ikx_a^2}{2}\left(\frac{1}{d} + \frac{1}{z}\right) - \frac{ik}{z}xx_a\right]. \tag{A3}$$

Under conditions of weak screening, the sought OV cores' positions are close to the beam axis, so one can assume

$$x \simeq y \simeq 0 \tag{A4}$$

and neglect the summands proportional to $x$, $x^2$ and $y$, $y^2$ when compared to the coordinate-independent terms. Further, for large $a$, the internal integral in (A2) can be estimated with the help of asymptotic formula valid for arbitrary function $f(x)$:

$$\int_a^{\infty} f(x) \exp(iKx^2) dx \simeq \frac{i}{2K} \frac{f(a)}{a} \exp(iKa^2) + O\left(\frac{1}{a^2}\right). \tag{A5}$$

Consequently,

$$\int_a^{\infty} dx_a\, (x_a + i\sigma y_a)^{|m|} P(x_a, x, z_c - iz_{Rc}) \simeq \frac{iP(a, 0, z_c - iz_{Rc})}{ak\left(\dfrac{1}{z_c - iz_{Rc}} + \dfrac{1}{z}\right)} (a + i\sigma y_a)^{|m|}, \tag{A6}$$

and the external integral of (A2) is estimated by the method of stationary phase which in couple with condition (A4) gives

$$\int_{-\infty}^{\infty} dy_a\, (a + i\sigma y_a)^{|m|} P(y_a, y, z_c - iz_{Rc}) \simeq \sqrt{\frac{2\pi i}{k\left(\dfrac{1}{z_c - iz_{Rc}} + \dfrac{1}{z}\right)}}\, a^{|m|}. \tag{A7}$$

Hence, we obtain the final representation for the integral term in the right-hand side of (A1):

$$-\frac{(-i)^{|m|+1}}{\sqrt{|m|!}} \sqrt{\frac{i}{2\pi}} \frac{k}{z} \left(\frac{z_{R0}}{z_c - iz_{Rc}}\right)^{|m|+1} \left[k\left(\frac{1}{z} + \frac{1}{z_c - iz_{Rc}}\right)\right]^{3/2} a^{|m|-1} P(a, 0, z_c - iz_{Rc}). \tag{A8}$$

The first term of (A1), with allowance for the near-axis condition (A4), reads

$$u^I(x,y,z) = u^{LG}(x,y,z_c+z) \simeq \frac{(-i)^{|m|+1}}{\sqrt{|m|!}} \left(\frac{z_{R0}}{z_c+z-iz_{Rc}}\right)^{|m|+1} \left(\frac{x+i\sigma y}{b_0}\right)^{|m|}. \tag{A9}$$

Now, gathering all terms of equation (A1), we find the asymptotic representation of the diffracted beam complex amplitude distribution (19), (20).

In case of the Kummer beam diffraction, in equation (A1) $u^I(x,y,z) = u^K(x,y,z_h+z)$, and due to (2) and [37] and the near-axis condition (A4),

$$u^K(x,y,z_h+z) \simeq \frac{z_e}{z+z_h}\sqrt{\frac{\pi}{2^{3|m|}}}(-i)^{|m|+1}\frac{z_R^{|m|+1}}{z_e^{|m|/2}(z_e-iz_R)^{|m|/2+1}}\left(\frac{x_e+i\sigma y_e}{b}\right)^{|m|} \tag{A10}$$

where

$$z_e = \frac{z+z_h}{1+(z+z_h)/R}, \quad x_e = x\frac{z_e}{z+z_h}, \quad y_e = y\frac{z_e}{z+z_h}.$$

In the integrand of (A1), we use the asymptotic expression for the Kummer beam amplitude (2) valid for $(x_a^2+y_a^2)/b^2 \gg 1$,

$$u_a(x_a,y_a) = \frac{z_{he}}{z_h}\left\{|m|(-i)^{|m|+1}\frac{z_{he}}{k(x_{ae}^2+y_{ae}^2)}\exp\left[\frac{ik}{2z_h}(x_a^2+y_a^2)\right]\right.$$

$$\left.+\frac{i}{2^{2|m|}}\frac{z_R^{2|m|+1}}{z_{he}^{|m|}(z_{he}-iz_R)^{|m|+1}}\exp\left[\frac{i}{2}\frac{k}{z_{he}-iz_R}(x_{ae}^2+y_{ae}^2)+\frac{ik}{2(z_h+R)}(x_a^2+y_a^2)\right]\right\}\left(\frac{x_a+i\sigma y_a}{\sqrt{x_a^2+y_a^2}}\right)^{|m|}. \tag{A11}$$

This expression is similar to the formal asymptotic of the Kummer function [41] modified with account for the non-zero wavefront curvature of the incident Gaussian beam [32]; however, in the considered range of $(x_a^2+y_a^2)/b^2 \simeq 10$, the formal asymptotic is not sufficiently accurate. To improve the approximation, the numerical coefficient in the second line of the asymptotic (A11) is modified, from $-2^{-|m|}i$ in [41], to $2^{-2|m|}i$ in (A11); validity of this correction was checked numerically.

Now both summands of (A11) should be substituted into the integral term of (A1). The first summand yields the corresponding summand of the integral term, which with omitted coordinate-independent coefficients obtains the representation (cf. expression (A2))

$$\exp\left[\frac{ik}{2z}(x^2+y^2)\right]\int_{-\infty}^{\infty}dy_a P(y_a,y,z_h)\int_a^{\infty}dx_a\frac{(x_a+i\sigma y_a)^{|m|}}{(x_a^2+y_a^2)^{|m|/2+1}}P(x_a,x,z_h) \tag{A12}$$

where function $P(x_a,x,z_h)$ is defined in (A3). Then, via the corresponding analogs of (A6) and (A7), we obtain

$$\int_a^{\infty}dx_a\frac{(x_a+i\sigma y_a)^{|m|}}{(x_a^2+y_a^2)^{|m|/2+1}}P(x_a,x,z_h) \simeq \frac{iP(a,0,z_h)}{ak\left(\frac{1}{z_h}+\frac{1}{z}\right)}\frac{(a+i\sigma y_a)^{|m|}}{(a^2+y_a^2)^{|m|/2+1}}, \tag{A13}$$

$$\int_{-\infty}^{\infty}dy_a\frac{(a+i\sigma y_a)^{|m|}}{(a^2+y_a^2)^{|m|/2+1}}P(y_a,y,z_h) \simeq \sqrt{\frac{2\pi i}{k\left(\frac{1}{z_h}+\frac{1}{z}\right)}}\frac{1}{a^2}, \tag{A14}$$

and after restoring the coordinate-independent coefficients, arrive at the total contribution of the first summand of (A11) to the integral term of (A1)

$$|m|(-i)^{|m|+1}\sqrt{\frac{i}{2\pi}}\frac{z_h}{z}\left[k\left(\frac{1}{z_h}+\frac{1}{z}\right)\right]^{-3/2}\frac{1}{a^3}P(a,0,z_h). \tag{A15}$$

Similar operations with the second summand of (A11) lead to the expression (cf. expressions (A2) and (A12))

$$\exp\left[\frac{ik}{2z}(x^2+y^2)\right]\int_{-\infty}^{\infty}dy_a P(y_a,y,z_d)\int_{a}^{\infty}dx_a\frac{(x_a+i\sigma y_a)^{|m|}}{(x_a^2+y_a^2)^{|m|/2}}P(x_a,x,z_d), \tag{A16}$$

where

$$\frac{1}{z_d}=\frac{1}{z_{he}-iz_R}\left(\frac{z_{he}}{z_h}\right)^2+\frac{1}{z_h+R}, \tag{A17}$$

which, finally, results in the following contribution to the integral term of (A1):

$$-\frac{ik}{2^{2|m|}}\sqrt{\frac{i}{2\pi}}\frac{z_{he}}{zz_h}\frac{z_R^{2|m|+1}}{z_{he}^{|m|}(z_{he}-iz_R)^{|m|+1}}\left[k\left(\frac{1}{z_d}+\frac{1}{z}\right)\right]^{-3/2}\frac{1}{a}P(a,0,z_d). \tag{A18}$$

Then, combining (A1), (A10), (A15) and (A18), we find the complex amplitude representation (7) – (11). Note that due to relation

$$\frac{1}{z_{he}-iz_R}\left(\frac{z_{he}}{z_h}\right)^2=\frac{i}{kb^2(z_h)}+\frac{1}{R(z_h)},$$

expressions (A17) and (A18) contain the radius $b(z_h)$ and the wavefront curvature radius $R(z_h)$ that the initial Gaussian beam, incident onto the VG, would have possessed in the screen plane if it had propagated "freely", without the VG-induced transformation.

**Post-publication note**

During the paper preparation, we met difficulties with numerical localization of one of the three OVs formed in the diffracted 3-charged Kummer beam at $z = 30$ cm (the "red" OV whose trajectory is presented in figure 2(e)). Namely, near $a = 2.35b$ the OV position appeared to be very sensitive to smallest shifts of the screen edge (see figure 6). At that moment, we could not carefully trace the OV evolution between $a = 2.35b$ and $a = 2.34b$, and the visible "jump" of the OV trajectory had been ascribed to the numerical inaccuracy. Based on the analogy with the trajectories presented in figures 2(c) and 2(d) where the distances between the cyan stars were "passed" quickly but some intermediate points still could be detected, we believed that for the red trajectory of figure 2(e), corresponding intermediate points also exist. As a result, the "solid" segment between the cyan stars in figure 2(e) had been obtained by interpolation along the arc-like "valley" that is well seen in lower parts of the figure 6 images.

However, a careful analysis performed after the paper has been published shows that there is a real discontinuity in the trajectory evolution, independent on the accuracy of calculations, and the curvilinear trajectory segment between the cyan arrows in figure 2(e) cannot be obtained in calculations. This "instantaneous transition" is illustrated by the straight dotted line in figure 2(e).

Interestingly, the possibility of "jumps" in the OV trajectories is dictated by very general considerations based on the simple model of the OV displacement described by equations (11) – (13). In the main text, we supposed that, in the near-axis region, the edge wave creates a spatially homogeneous contribution (11) – in fact, the terms containing $x$ and $y$ were discarded. These can be easily restored, taking into account the slightly inclined wavefront of the edge wave in the ($xz$) plane. With this improvement, and allowing for (6), expression (11) will read

$$E_{\text{edge}} = D_0(a) \exp\left[ik\left(z + \frac{a^2}{2z} - x\frac{a}{z}\right)\right] = D_0(a) \exp\left[ik\left(z + \frac{a^2}{2z} - r\frac{a}{z}\cos\phi\right)\right]. \quad (23)$$

Accordingly, the equation for the OV core azimuth (13) is modified to

$$\phi + \frac{kra}{mz}\cos\phi = \text{const} + k\frac{a^2}{2mz}. \tag{24}$$

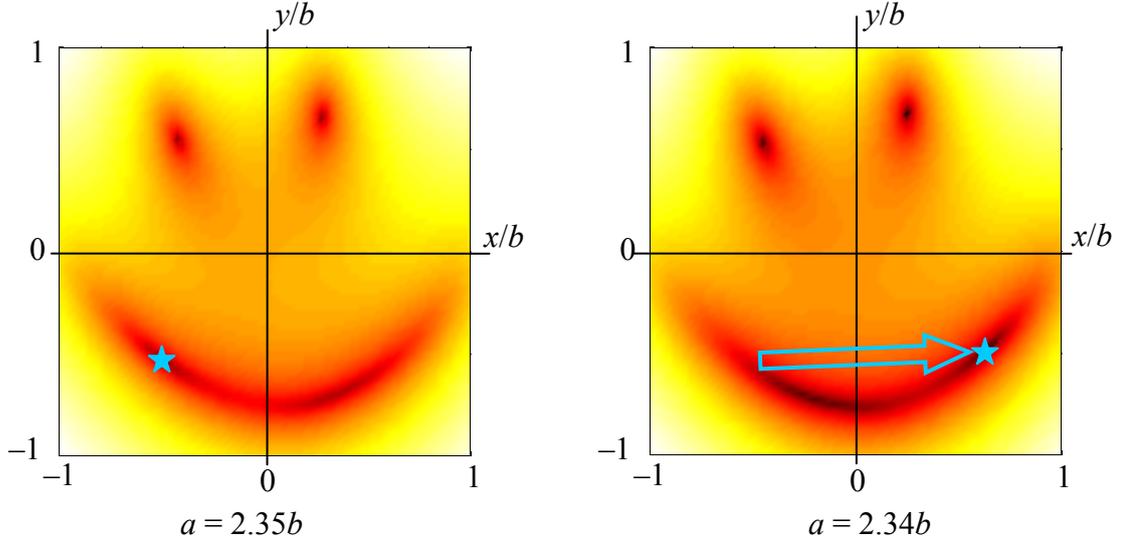

**Figure 6**. Pseudocolor maps of the near-axis region for the diffracted Kummer beam with topological charge $m = -1$ for the screen edge positions $a = 2.35b$ and $a = 2.34b$ (cf. figure 2 (c) – (e)). The images represent the transformed intensity distribution $\left[|u(x,y)|^2\right]^{1/15}$ with enhanced visibility of the amplitude zeros [35]; dark spots in the upper parts of the images show the practically unchanged positions of the 'blue' and 'black' OVs whose trajectories are presented in figures 2(c), (d). In contrast, the third OV core marked by cyan star performs a "jump" shown by the arrow.

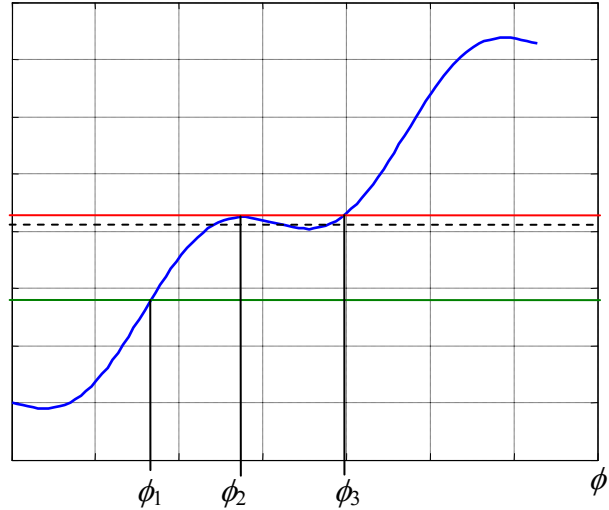

**Figure 7**. Illustration of the solution of equation (24). The blue curve is a plot of the left-hand side expression of (24) at $|kra/mz| = 1.4$, horizontal lines symbolize the different right-hand side values. The green line shows the one-to-one correspondence between $\phi$ and $a$ that is realized for a monotonous function in the left-hand side and leads to a continuous OV core trajectory with monotonously varying azimuth $\phi_1$; the red line describes the conditions of a "jump" from $\phi_2$ to $\phi_3$ upon a smallest variation of $a$. The black dashed line corresponds to the hypothetic case when additional OVs can exist.

Now the OV core angular coordinate is determined by the transcendent equation which makes it difficult to express an explicit solution to the problem. However, a qualitative analysis is obvious (see figure 7). Due to the trigonometric term, the left-hand side of (24) can be non-monotonic, and this is the reason for possible jumps in the $\phi(a)$ dependence. One can easily verify that the monotonic character of the right-hand function is violated when

$$\left|\frac{kra}{mz}\right| > 1,$$

which can occur even in the near-axis region ($r \ll b$) for large enough $a$ and not very high $z$. In our present conditions of $k \approx 10^5$ cm, $b = 0.232$ mm (equations (5)), $z = 30$ cm, $a = 2.35b$, $r \approx 0.72b$ (figure 6) one finds $|kra/mz| \approx 1.0$. This shows that the "jump" in the OV-core trajectory observed in figure 2(e) is not striking but an expected phenomenon.

Probably, this effect can be incorporated into the frame of asymptotic approach developed in the Appendix; to this end, one should loose the limitations of (A4) and take some terms proportional to $x$ and $y$ into account. This will destroy the analytical transparency of the asymptotic formulae and make them less suitable for calculations but can be useful for revelation of fine qualitative features of the OV trajectories. In particular, one may suspect that the "fast" segments of the "blue" and "black" trajectories between the cyan stars in figures 2(c) and (d) are not completely continuous and also contain some "jumps". Also, the dashed horizontal line in figure 7 suggests that three amplitude zeros can emerge instead of a single secondary OV – is this situation really observable or not, is currently unclear. All such questions require additional examination of the problem that will be made elsewhere.